# Predicting Mechanical Properties of Lignin-Containing Polyurethane Rigid Foams from Microstructure Using Convolutional Neural Networks

Ilige S. Hage[a*], Charbel Y. Seif[b], Jose Enrico Q. Quinsaat[c], Daniel J. van de Pas[c], Richard Vendamme[d], Walter Eevers[d,e], Karolien Vanbroekhoven[d], Elias Feghali[d, f*]

[a*] Mechanical Engineering Department, Notre Dame University-Louaize, P.O. Box: 72, Zouk Mosbeh, Zouk Mikael, Lebanon.

[b] Mechanical Engineering Department, American University of Beirut, P.O. Box 11-0236, Riad El-Solh, Beirut 1107 2020, Lebanon

[c] Scion Group, Bioeconomy Science Institute, Titokorangi Drive, Private Bag 3020, Rotorua 3046, New Zealand

[d] Sustainable Polymer Technologies (SPOT) Team, Flemish Institute for Technological Research (VITO), Boeretang 200, Mol, Belgium.

[e] Department of Chemistry, University of Antwerp, Groenenborgerlaan 171, 2020 Antwerp, Belgium

[f] Chemical Engineering Department, Notre Dame University-Louaize, P.O. Box: 72, Zouk Mosbeh, Zouk Mikael, Lebanon.

[*]Corresponding Author (ilige.hage@ndu.edu.lb)

## Abstract

**Bio-based alternatives for conventional rigid foams have proven to be good substituents owing to their enhanced sustainability and competitive performance. However, because their manufacturing processes are complex and destructive testing is often impractical, this study investigates whether microstructural features can be correlated with mechanical properties in lignin-containing rigid polyurethane (PU) foams using machine-learning approaches. Various types and percentages of lignin-based polyols were investigated as a partial polyol replacement. Scanning electron microscopy (SEM) images and corresponding mechanical compression data were used to train a custom state-of-art dual-head convolutional neural network (CNN) targeting the specific prediction of density, specific compression modulus, specific yield strength, and specific compression strength. The CNN was optimized with a weighted multi-output loss function, achieving strong predictive performance $R^2$ values ranging from 0.850 to 0.91 and correlation coefficients above 0.92 while maintaining mean absolute error percentages below ≈9%. This proves the trained network capability to predict and capture morphological features governing load bearing responses. On the other hand, Grad-CAM visualization revealed that the network focused its predictions on physically**

**meaningful microstructural regions such as cell walls and strut junctions, which confirm that the proposed network can be classified as an interpretable, non-destructive and data-driven framework for predicting and understanding bio-based PU foams mechanical behavior, hence reducing the inconvenience caused by time consuming manufacturing and destructive testing.**

## 1. Introduction

Insulation, mechanical resilience and lightweight are essential materials properties within several industrial sectors including transportation and construction. Polyurethane (PU) rigid foams are characterized by such properties as proven in Kausar [1] and Mistry et al. [2]. However, conventional PU foams have drawbacks related to environmental/health impact and sustainability concerns since they are manufactured using petroleum-based polyols and polyisocyanates. On the other side, lignin is the most abundant biobased polymer derived from biomass, which can be used as a polyol as proven in Tran et al. [3]and Ma et al. [4] to potentially substitute polyols in PU formulations. PU foam's morphology and mechanical response are highly affected by lignin's heterogeneity and limited reactivity as stated in Gao et al. [5]. Moreover, lignin is also known for its flame retardancy properties (Yang et al. [6]) which makes it a great candidate for polyols replacement to minimize the potential fire risk during the foam production and the corresponding product service.

PU foams microstructure specifically the size, distribution and shape of foam cells affect their mechanical properties such as compressive strength modulus and density as described by Hawkins et al. [7]. In Bhagavathula et al. [8], it was shown the density, microstructure and strain rate jointly affect the compressive behavior of polymeric foams. Characterizations done on PU foams using destructive mechanical testing methods which are time consuming due to the tedious preparation of test samples. However recent literature focuses on the use of artificial intelligence in predicting properties based on imaging data which was proven to be efficient for porous material structurally similar to PU foam.

In Omid et al. [9] Goodarzi and Bahramian [10], convolution neural networks (CNNs) were proven as a successful tool in porous media with foam-like features to predict porosity, young's modulus and compressive strength. Liu et al. [11] trained a CNN on microstructural images of porous, foam-like ceramic media to predict effective stiffness with high accuracy. The study in Yin et al. [12] applied image-based deep learning to foam ceramics to infer pore structure characteristics directly from microstructural images, demonstrating that porosity related descriptors can be estimated from imaging alone.

New avenues for material characterization are now achievable due to digital image processing with machine learning advancements. In particular, CNNs can capture complex morphological patterns from microscopic images and predict mechanical properties without the need for physical destructive testing. In Giannis et al. [13] a 3D-CNN model is proposed to reconstruct 3D shapes of granular particles from multiple 2D sections via training on synthetic particles with known geometric properties. Low error rates in predicting key shape descriptors were achieved using this model, this reduces the need for X-ray microtomography and potential applications in materials science. In Sun et al. [14] a fully CNN based on StressNet is modified to predict stress fields in fiber-reinforced composites from tomography images using data from non-linear finite element simulations. The model was able to accurately capture stress distributions especially around fibers and makes predictions in seconds instead of 92.5 hours for traditional simulations, demonstrating the potential of Machine learning (ML) for rapid structural analysis and damage site identification. An adaptive residual CNN is proposed in Song et al. [15] to predict material properties from microstructure images. The network was structured to be optimized based on prediction error using simulated annealing and it was tested on SEM images of polymer explosives. The model achieved an MAPE of 1.3% and $R^2$ of 0.943. In this study Grad-CAM was used to visualize microstructure-property correlations, highlighting the method's effectiveness and interpretability. In Shin et al. [16] a CNN-based model was introduced to predict mechanical behavior of monocrystalline graphene, accounting for complex grain boundaries and defects. The training robustness was enhanced using data augmentation via periodic boundary condition, also principal component analysis was utilized to reduce data dimensionality and to boost learning efficiency. The model achieved high accuracy in generating realistic atomics structures, and it showed effectiveness in handling interpolation and extrapolation. Grad-CAM was also utilized to confirm the model's ability to identify critical structural feature influencing mechanical properties.

Artificially intelligent image-based, nondestructive approaches have been applied to several porous materials, yet few studies have explored lignin-modified PU rigid foams. A research gap persists in applying such artificial intelligence predictive methods to bio-based, variable structure PU systems specifically lignin modified foams. In Hage et al. [17], neural networks were utilized to classify lignin-based PU foams, based on microscopic imaging of their respective compositions, and to explore their structure-property relationships. Walicki et al. [18] demonstrated the effectiveness of usage of statistics, machine learning and deep learning methods in examining microstructures of foam polymers and prediction their macroscopic properties. This work presents a dual head CNN architecture for the prediction of mechanical properties and density of PU rigid foams that are lignin-modified from their microscopic cell structure. This novel network architecture is designed to predict multiple parameters by grouping those with similar response behaviors within identical network branches. The main objective is to develop a neural network to specifically predict properties of density, specific compression modulus, yield stress and compressive strength from micro images, with a target to investigate the effect of variation in lignin content on foam morphology and mechanical performance and to validate a multi-output dual head CNN. To the best of the authors' knowledge, this is the first time microscopic imaging has been applied in the development of a CNN to predict mechanical behavior in lignin-modified PU foams. The method is a rapid, non-destructive alternative to traditional mechanical testing, and provides valuable information about the influence of natural additives like lignin on microstructure property relationships, advancing the design of sustainable bio-based materials to render robust and reliable materials solutions. This method can help material engineering to predict and assign the needed structural component of lignin in PU foam-based polymers based on desired mechanical properties.

## *2. Experimental Setup*

Compressive testing of the PU foams was performed on a universal testing machine, and microscopic images were captured at 40X magnification using SEM. A total of 107 images were utilized in this work after discarding blurred or out of alignment images. PU foams were produced using polymeric methylene diisocyanate (pMDI), sorbitol and sucrose-based polyether polyols, silicon surfactant, blowing agent and tertiary amine catalyst (ref). Dihydroconiferyl alcohol (DCA) was prepared from eugenol and according to the protocol described in Driscoll et al. [19]. Lignin-based polyols like Kraft lignin (KL) and lignin hydrogenolysis oil (LHO) from *Pinus radiata*. were used to replace sorbitol polyols with an isocyanate index of 1.3, with up to 50 wt.% (Quinsaat et. al [20] Feghali et al. [21]). ASTM D1621-10 standard was utilized for compression testing with 15x10x10 mm specimens at 23°C and 50% humidity for 48 hours and at a strain rate of 10% per minute to 80% compression. Sections of 5x5 mm were images using SEM with chromium coating utilizing JEOL JSM-6700F microscope at 3 kV. Figure 1 shows some Representative SEM images for 25 wt.% lignin foams (Quinsaat et. al [20]) along with each image corresponding experimental measurements of density ($g \cdot cm^{-3}$), specific compression modulus ($MPa/(g \cdot cm^{-3})$), specific yield strength ($MPa/(g \cdot cm^{-3})$) and specific compression strength ($MPa/(g \cdot cm^{-3})$).

| **Sample** | | **Microscopic Image (40×Uncompressed)** |
|---|---|---|
| **CF** | | |
| **Density** | 0.067 | |
| **Specific Compression Modulus** | 117.5 | |
| **Specific Yield Strength** | 4.18 | |
| **Specific Compression Strength** | 5.99 | |
| **LHO** | | |
| **Density** | 0.063 | |
| **Specific Compression Modulus** | 115.63 | |
| **Specific Yield Strength** | 4.22 | |
| **Specific Compression Strength** | 5.9 | |

KL

| Density | 0.049 |
| --- | --- |
| Specific Compression Modulus | 138.61 |
| Specific Yield Strength | 4.3 |
| Specific Compression Strength | 6.4 |

LHO-MD

| Density | 0.0634 |
| --- | --- |
| Specific Compression Modulus | 202.5 |
| Specific Yield Strength | 7.524 |
| Specific Compression Strength | 7.31 |

LHO-O

| Density | 0.0606 |
| --- | --- |
| Specific Compression Modulus | 176.02 |
| Specific Yield Strength | 4.67 |
| Specific Compression Strength | 7.73 |

DCA-D

| Density | 0.0664 |
| --- | --- |
| Specific Compression Modulus | 87.97 |
| Specific Yield Strength | 2.88 |
| Specific Compression Strength | 3.033 |

| DCA | |
|---|---|
| **Density** | 0.0457 |
| **Specific Compression Modulus** | 155.62 |
| **Specific Yield Strength** | 6.11 |
| **Specific Compression Strength** | 4.59 |

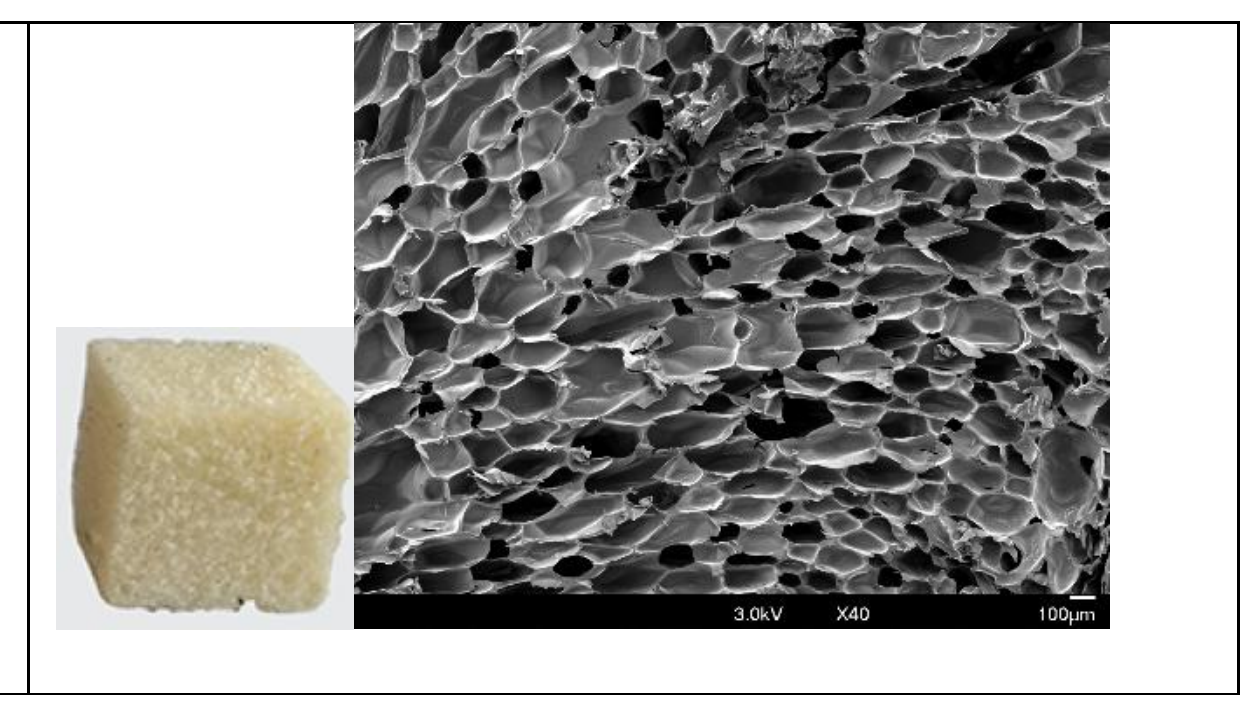

**Figure 1: Images of PU foams prepared using the different lignin-based polyols at a loading of 25 wt.% of the polyol component and their SEM images, before compression at 40X magnification *(Source: Reproduced from Ref.* [20] *with permission from the American Chemical society)* associated with their corresponding properties of density (g·cm⁻³), specific compression modulus (MPa/( g·cm⁻³)), specific yield strength (MPa/( g·cm⁻³)) and specific compression strength (MPa/( g·cm⁻³)) for each foam**

## 3. Materials and Methods

### 3.1 Data Acquisition and Preprocessing

Based on the compression testing done in Quinsaat et. [20]. The mechanical ground-truth data of density and specific compression modulus, yield strength and compression strength comprised of four properties was modeled as follows:

$$y = [\rho, E_c, \sigma_y, \sigma_c]^\top \quad (1)$$

where:

- $\rho$= density ($g.cm^{-3}$),
- $E_c$= specific compressive modulus ($MPa/(g.cm^{-3})$),
- $\sigma_y$= specific yield stress ($MPa/(g.cm^{-3})$),
- $\sigma_c$= specific compressive strength ($MPa/(g.cm^{-3})$).

Descriptive statistics and Pearson correlation analysis confirmed that the density (ρ) is weaky correlated with the mechanical properties, whereas the compressive modulus $E_c$, yield stress $\sigma_y$ , and compressive strength $\sigma_c$ exhibit good to strong mutual correlation (r=0.61-0.93). This motivated the design of a dual-head CNN, where one output head predicts the density independently, and the second multi-output head jointly predicts the 3 correlated mechanical parameters. This is demonstrated in figure 2 that shows a color-coded correlation heat map illustrating the pairwise Pearson coefficients among the 4 measured properties. The diagonal cells (value=1) represent perfect self-correlation while off-diagonal entries quantify linear dependencies between parameters. The results indicate that density shows only weak correlation with the mechanical parameters (r=0.29-0.36), suggesting that the density vary largely independently from the stress-strain response of the foams. On the contrary, the 3 mechanical parameters demonstrate strong independence with r=0.61-0.93, implying that stiffness and strength in lignin-based rigid PU foams scale almost proportionally once polymer morphology and crosslink density are fixed. This separation between a low-correlation density domain and a high-correlation mechanical domain based on Cohen [22], which supports the two-head CNN structure adopted in this work.

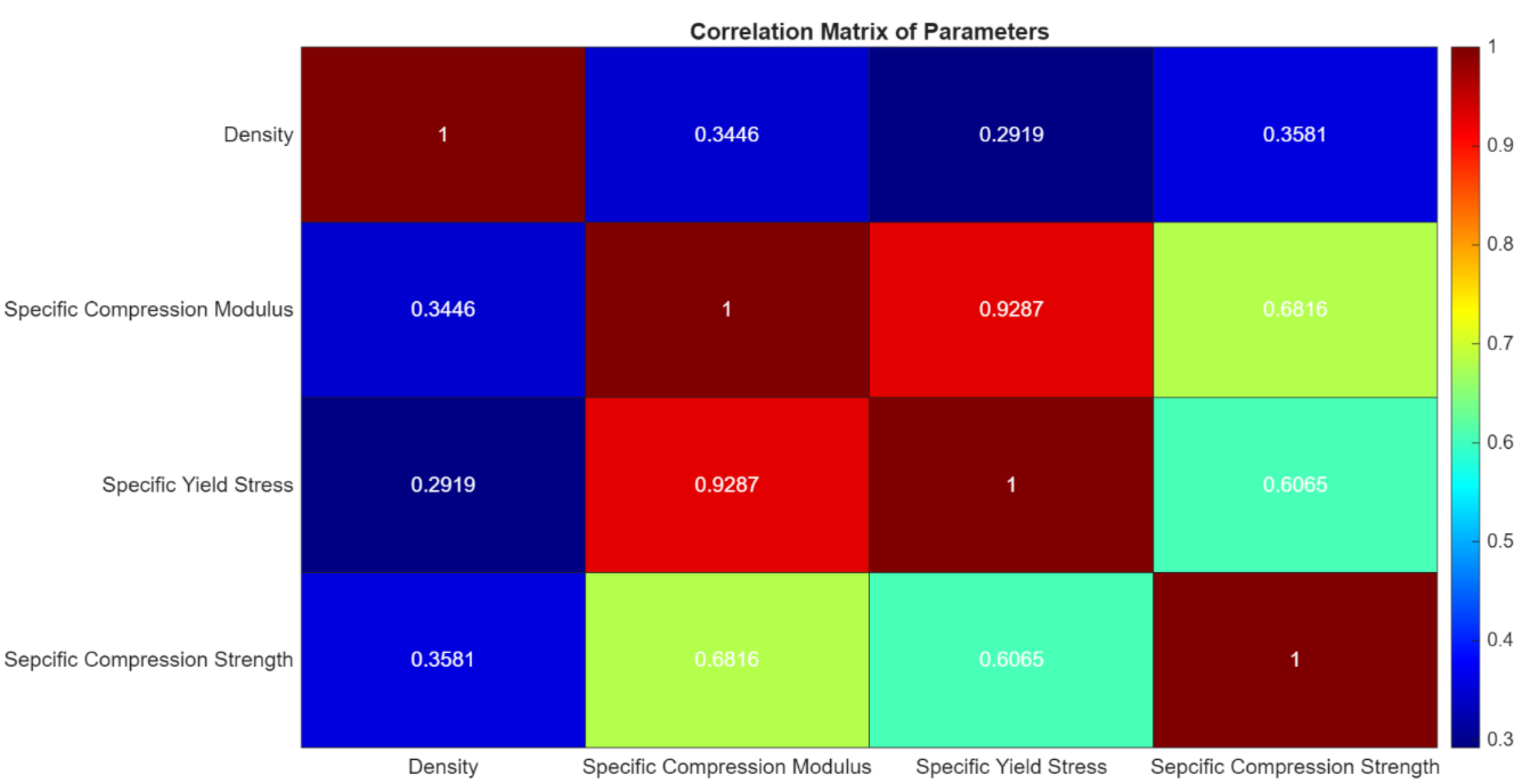

**Figure 2: color coded heat map for the correlation matrix of the density and the mechanical properties**

Figure 3 completes the heat-map analysis by displaying pairwise scatter plots among all variables with histograms on the diagonal showing their marginal distributions. A roughly normal distribution is shown for the density centered near 0.06 g.cm$^{-3}$, while the mechanical properties span broader ranges, reflecting the diversity of polymer network architectures and lignin substitution ratios. The scatter plots reveal nearly linear relationships between $E_c$, $\sigma_y$, and $\sigma_c$ confirming the strong correlation shown in the heat map in figure 2. In contrast, density points are more dispersed and show only weak trends with the mechanical properties, indicating that variations in foam porosity and cell geometry do not directly yield proportional and linear changes in stiffness or strength.

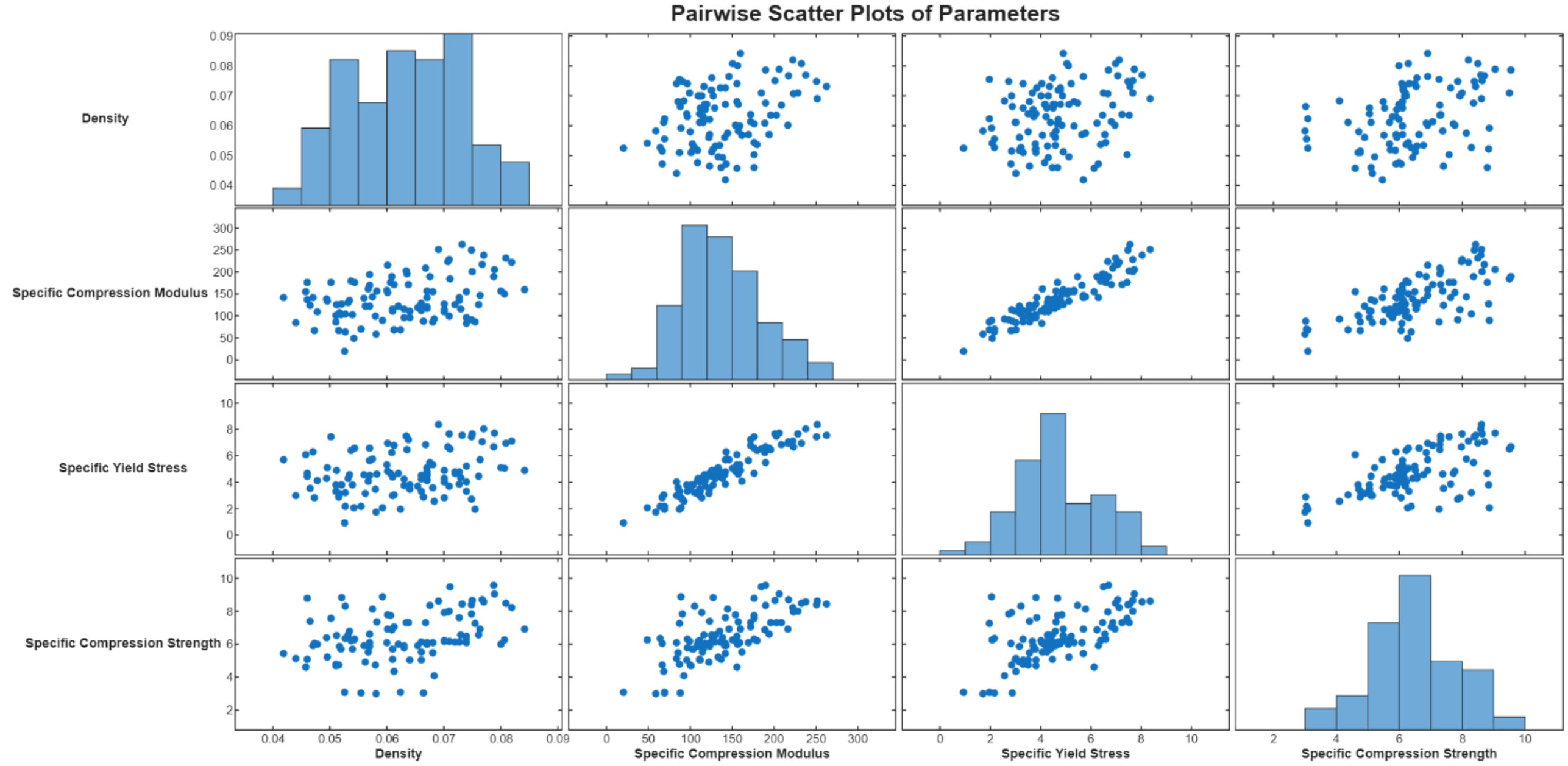

**Figure 3: Pairwise Scatter plots correlating parameters of density, compression modulus, yield stress and compression strength.**

Figures 2 and 3 highlight the natural clustering of mechanical responses versus density, justifying their treatment as 2 correlated but distinct prediction domains. Accordingly, a dual-head CNN was implemented with one head specialized in predicting density and the other dedicated to predicting the 3 mechanical properties.

This proposed network will utilize SEM images which were resized to 240x320x3 pixels and normalized to [0, 1]. Target labels were standardized using Z-score normalization as reported in Gareth et al. [23]:

$$\tilde{y}_i = \frac{y_i - \mu_i}{\sigma_i} \quad (2)$$

where $\mu_i$and $\sigma_i$are the mean and standard deviation of the i-th property.

The normalized dataset thus formed the final regression label matrix:

$$\tilde{Y} = \begin{bmatrix} \tilde{\rho}_1 & \tilde{E}_{c,1} & \tilde{\sigma}_{y,1} & \tilde{\sigma}_{c,1} \\ \vdots & \vdots & \vdots & \vdots \\ \tilde{\rho}_N & \tilde{E}_{c,N} & \tilde{\sigma}_{y,N} & \tilde{\sigma}_{c,N} \end{bmatrix} \in \mathbb{R}^{N\times 4} \quad (3)$$

Input images are converted into 4D tensors [H×W×C×N] or the deep network, and labels are reshaped into [1×batch] for density and [3×batch] for the mechanical properties. A routine was created to  ensure alignment between labels and input samples across training, validation, and test datasets. In addition to normalization; minimum, maximum, and standard deviations were computed for each property to monitor variance preservation within the training data sets 40%, testing 30% and validation datasets 30% across foam types following equation (4):

$$y_{k,\min} = \min_i y_{i,k}, \quad y_{k,\max} = \max_i y_{i,k}, \quad \sigma_k^{norm} = \mathrm{std}(\tilde{y}_{:,k}) \qquad (4)$$

### 3.2 Dataset Partitioning and Augmentation

To improve model generalization and increase dataset diversity under limited sample availability (~107 images), a controlled stochastic augmentation scheme was applied using MATLAB®. Augmentations were restricted to small geometric transformations to preserve foam microstructural fidelity (cell size and orientation). Each mini-batch $X_i$was randomly perturbed by a transformation $T_i$, forming an augmented dataset:

$$X_i' = T_i(X_i) \qquad (5)$$

where the transformation operator $T_i$is drawn from the augmentation distribution:

$$T_i \sim \mathcal{A}(\Delta\theta, s, t)(6)$$

with:

$$\begin{array}{llc} \Delta\theta & \in [-5^\circ, 5^\circ] & \text{(random rotation)} \\ s & \in [0.97, 1.03] & \text{(isotropic scaling)} \\ t_x, t_y & \in [-3,3] \text{ pixels} & \text{(random translation)} \\ r_x & \in \{0,1\} & \text{(random horizontal reflection)} \end{array} \qquad (7)$$

The transformations were applied only to training batches ($X_{train}$), leaving validation and test subsets unmodified to ensure consistent evaluation conditions.

This approach can be formally expressed as:

$$\mathcal{D}_{train}' = \{(T_i(X_i), \tilde{y}_i) \mid T_i \sim \mathcal{A}\}, \mathcal{D}_{val,test}' = \{(X_i, \tilde{y}_i)\}. \qquad (8)$$

To help the CNN learn rotation-invariant and translation-invariant microstructural representations, small angle rotations and subpixel translations were performed in line with Shorten et al. [24], Wang et al. [25]. A mild scaling perturbation (±3%) simulated possible variations in microscopic magnification or focus. This augmentation balances between statistical enrichments and physical plausibility, ensuring that this artificial transformation does not distort pore geometry or cell-wall thickness distribution. To ensure this consistency between SEM images and their corresponding targets, a custom MATLAB® datastore was implemented. After augmentation, batching or shuffling operations the code checks to maintain synchronized image-label pairs, where each

image–label pair was denoted as $(X_i, \tilde{y}_i)$, where $X_i \in \mathbb{R}^{240\times320\times3}$and $\tilde{y}_i \in \mathbb{R}^4$. The custom datastore $\mathcal{D}$is defined as:

$$\mathcal{D} = \{(X_i, \tilde{y}_i) \mid i = 1, \dots, N\}. \qquad (9)$$

Internally, each batch $\mathcal{B}_j = \{(X_i, \tilde{y}_i)\}_{i\in\mathcal{I}_j}$is generated using a mini-batch queue, where the function applies a preprocessing pipeline ( as shown in equation 10) ensuring image normalization, resizing, and label integrity are preserved:

$$(X_i, \tilde{y}_i) \overset{\text{augmenter}}{\rightarrow} (T_i(X_i), \tilde{y}_i) \overset{\text{preprocess}}{\rightarrow} (\bar{X}_i, \bar{y}_i) \quad (10)$$

The augmentation distribution $T_i \sim \mathcal{A}(\Delta\theta, s, t)$ was defined in Eq. (6) .

This work proposed a novel dual-head convolutional neural network that was designed to predict the density and the mechanical properties form lignin PU foams images. The network follows a shared-backbone and multi output-architecture that enables joint learning of correlated physical properties. The network architecture is shown in Figure 4 and Table 1 lists the details of each part of the network. Its backbone consists of three convolutional stages, each composed of a 3×3 convolution layer (with He initialization), batch normalization, LeakyReLU activation (α = 0.2), and 2×2 max pooling. The number of filters increases progressively (2×, 4×, and 8× netbase), capturing morphological hierarchies from low-level textures (cell size, edges) to higher-level patterns (pore connectivity and wall thickness). A Global Average Pooling (GAP) layer replaces conventional flattening, generating a compact 256-dimensional feature descriptor that is spatially invariant. This is followed by a shared fully connected layer of 4×netbase neurons with LeakyReLU (α = 0.1) activation and a 30% dropout, forming a 128-dimensional latent embedding distributed to both regression heads.

The Density head is formed of three dense layers (128, 128 and 64 neurons) with LeakyReLU activations, and 20% dropout comprises the density head, which leads to a single linear output neuron. For the mechanical properties head, the same hierarchy was followed but ends with a three-neuron output layer that represented the 3 target mechanical properties. Bias regularization and He initialization to ensure stable convergence were employed by all layers. To allow robust learning from small microscopy datasets, this architecture effectively balances expressive power and regularization. The share backbone supports feature reuse across correlated outputs, while other branches were tasked specifically to enhance accuracy for each target. GAP, dropout and LeakyReLU improved generalization and convergence stability. The proposed network provides a compact and powerful framework for multi-property prediction from microscopic images, enabling extraction of physically meaningful relationships between foam morphology and mechanical behavior.

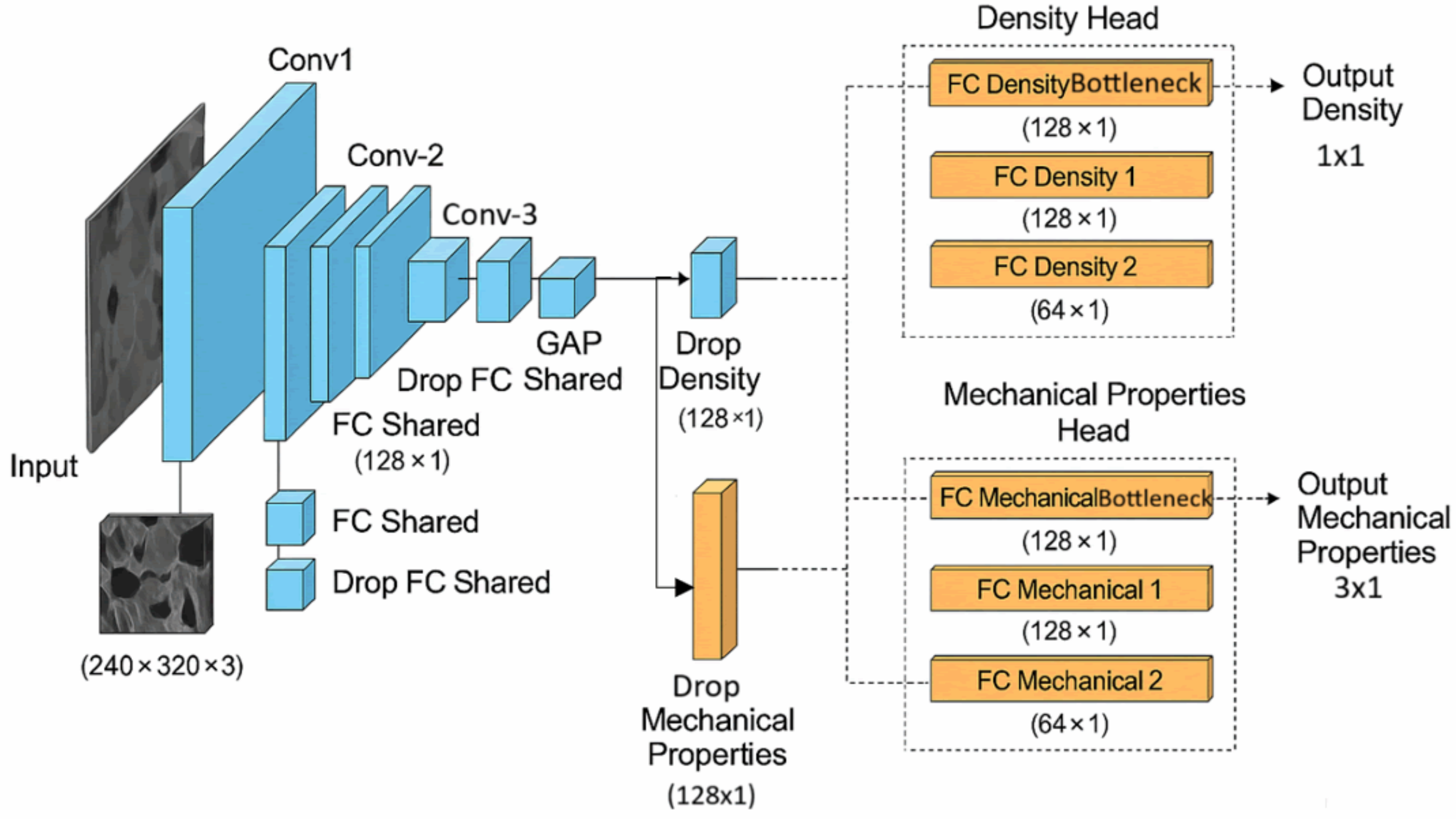

**Figure 4: Dual head network architecture**

**Table 1: The dual head network architecture detailed structure**

| Section | Layer Name | Size |
|---|---|---|
| **Backbone** | Input | [240 × 320 × 3] |
| | Conv1 | [240 × 320 × 64] |
| | Pool1 | [120 × 160 × 64] |
| | Conv2 | [120 × 160 × 128] |
| | Pool2 | [60 × 80 × 128] |
| | Conv3 | [60 × 80 × 256] |
| | Pool3 | [30 × 40 × 256] |
| | Global Average Pooling (GAP) | [1 × 1 × 256] |
| | Fully Connected (FC Shared) | [128 × 1] |
| | Dropout (Drop FC Shared) | [128 × 1] |
| **Density Head** | FC Density Bottleneck | [128 × 1] |
| | FC Density 1 | [128 × 1] |
| | FC Density 2 | [64 × 1] |
| | Output Density | [1 × 1] |
| **Mechanical Properties Head** | FC Mechanical Bottleneck | [128 × 1] |

| | FC Mechanical 1 | [128 × 1] |
|---|---|---|
| | FC Mechanical 2 | [64 × 1] |
| | Output Mechanical Properties | [3 × 1] |

### 3.4 Training and Optimization

In this work, the network was trained by using stochastic gradient descent with momentum (SGDM) as the main optimizer. Data is divided into 40% training set, 30% testing set and 30% validation set, each set is selected to have a mean and standard deviation identical to the original data.

The training settings used a momentum coefficient of β=0.9 to stabilize the convergence and accelerate the movement in relevant gradient directions, an initial learning rate of η=2×10-3, and a total of 200 epochs. During training, mini-batches of size 12 were randomly drawn from the training set and scrambled at each epoch to increase convergence stability and generalization. L2 regularization (λ=1×10-5) and gradient clipping (‖∇θL‖≤1) are applied to avoid overfitting and prevent exploding gradients. According to Ruder [26]:, the update rule for the parameters of the SGDM is:

$$v_t = \beta v_{t-1} + (1-\beta)\nabla_\theta L_t, with\ \theta_{t+1} = \theta_t - \eta v_t \ (21)$$

where $v_t$ is the velocity term, β is the momentum constant, $\nabla_\theta L_t$ is the gradient of the loss function with respect to the network parameters, and η is the learning rate.

Gradients are selectively backward propagated to the network parameters based on layer grouping. The backbone receives gradients from all tasks, while head layers receive gradients only from their corresponding output. This masking of gradients ensures that there is no interference between task-specific features. Optionally, gradients can be globally clipped to avoid exploding gradients. Validation was performed every 10 iterations using a dedicated validation dataset with a patience of 20 epochs. The model corresponding to the lowest validation loss was preserved as the best network. Data is loaded via mini-batch queues which invoked a preprocessing function efficiently normalizing and augmenting the input. Loss terms were tracked at both iteration and epoch levels; specific weighting was applied to different task heads during multi-output learning to deal with the task imbalance issue ($[w_{density}, w_{others}]$=[1.4,0.4]).

Along with SGDM, the optimizer AdamW (Adam [27], I. Loshchilov and F. Hutter [28]) was utilized , for adaptive weight update. AdamW combines momentum-based gradient accumulation with per-parameter learning rate adaptation along with decoupled weight decay. Update equations for it are defined as:

$$m_t = \beta_1 m_{t-1} + (1-\beta_1)\nabla_\theta L_t, v_t = \beta_2 v_{t-1} + (1-\beta_2)(\nabla_\theta L_t)^2$$

$$\hat{m}_t = \frac{m_t}{1-\beta_1^t}, \hat{v}_t = \frac{v_t}{1-\beta_2^t} \quad (22)$$

$$\theta_{t+1} = \theta_t - \eta \frac{\hat{m}_t}{\sqrt{\hat{v}_t} + \epsilon} - \eta\lambda\theta_t$$

where $m_t$ and $v_t$ denote the first and second moment estimates of the gradient, respectively; $\beta_1$=0.9 and $\beta_2$=0.999 are exponential decay rates; $\epsilon = 10^{-8}$ ensures numerical stability; and $\lambda=10^{-4}$ represents the weight decay coefficient.

The learning rate for the backbone network was initialized at $2\times10^{-4}$, while the head networks used a reduced multiplier of 0.75. In addition, for balanced learning between density and auxiliary outputs, EMA smoothing with a factor of $\tau$=0.9 is applied to hybrid loss components to dynamically adapt task weights during training.

### 3.4.1 Adaptive Multi-Phase Training and Stability Control

The training process was organized in three adaptive phases designed to ensure stable convergence and optimal utilization of pre-trained backbone features. In Phase 1 (Heads-Only Warmup), only the task-specific output heads were updated, while the backbone layers were frozen using a minimal learning rate ($\eta_{backbone}=10^{-6}$). This warm-up strategy prevents degradation of pretrained representations and allows rapid adaptation of high-level task mappings (Bengio [29]). A cosine annealing learning rate schedule (I. Loshchilov and F. Hutter[30]) was applied to the head layers to progressively decrease the step size and enhance generalization performance:

$$\eta_t = \eta_{min} + \frac{1}{2}(\eta_0 - \eta_{min})\left(1 + \cos\left(\frac{\pi t}{T}\right)\right) \quad (23)$$

where $\eta_0$ and $\eta_{min}$ represent the initial and minimum learning rates, respectively, t is the current epoch, and T is the total number of epochs in the phase.

The backbone was gradually unfrozen during phase 2 (fine-tuning) , and its learning rate was increased according to a controlled multiplicative ramp-up allowing deeper convolutional filters to adapt without causing instability:

$$\eta_{\text{backbone}}^{(k+1)} = min\left(1.15\eta_{\text{backbone}}^{(k)}, 2\times10^{-5}\right) \quad (24)$$

The final phase (stabilization) used a reduced learning rate ($\eta_{backbone}\leq3\times10^{-5}$) to refine weights and stabilize predictions in later training stages.

Gradient clipping was employed at both the group and elementwise levels to maintain robust optimization:

$$\tilde{g} = \text{clip}(g, \|g\|_F \leq c_N, |g| \leq c_V) \quad (25)$$

where g denotes the gradient tensor, ‖g‖F its Frobenius norm, and $c_N$,$c_V$ the norm and elementwise clipping thresholds. The backbone used stricter limits ($c_N$=5, $c_V$=0.1) than the output heads ($c_N$=50, $c_V$=1.0) to avoid disruptive updates in low-level feature extractors.

A mechanism of dynamics task reweighting was also integrated to balance learning between the density and auxiliary outputs. Task weights were updated according to the inverse of the exponentially smoothed loss magnitudes:

$$w_i = \frac{1/\mathrm{EMA}(L_i)}{\sum_j 1/\mathrm{EMA}(L_j)}, \mathrm{EMA}(L_i) = \tau\mathrm{EMA}(L_i) + (1-\tau)L_i \qquad (26)$$

where τ=0.9is the e smoothing constant. This adaptive weighting ensures that slower-converging tasks receive proportionally higher emphasis, preventing overfitting to easier objectives Chen et al. [31].

Dynamic weighting strategies are involved in the training process as follows:

- Plateau detection: If training stagnates, correlation loss weight is reduced.
- Clamping and penalty strengths schedule over epochs in warmup and fine-tuning phases:
- Learned uncertainty weighting: this option primarily scales the losses with respect to predicted confidence.

These mechanisms improve the convergence stability and balance multiple tasks during multi-task learning.

To evaluate predictive performance across both outputs, Pearson correlation coefficients (defined by Benesty et al. [32]) were calculated between predicted ($\hat{y}_i$) and true target values ($y_i$), providing a quantitative measure of linear agreement between model outputs and ground truth:

$$r = \frac{\sum_i (y_i - \bar{y})(\hat{y}_i - \bar{\hat{y}})}{\sqrt{\sum_i (y_i - \bar{y})^2 \sum_i (\hat{y}_i - \bar{\hat{y}})^2}} \quad (27)$$

### 3.4.2 Early Stopping, Plateau Adaptation, and Model Evaluation

Early Stopping was employed during training to avoid overfitting and unnecessary computation, where the validation loss $L_{val}$ was monitored at each epoch, and the best-performing model fθ was saved whenever a significant improvement was detected, defined as

$$L_{\mathrm{val}}^{(t)} < L_{\mathrm{best}} - \Delta_{min} \quad (28)$$

where $\Delta_{min}=10^{-3}$ represents the minimum meaningful improvement threshold. If no improvement occurred for p=20 consecutive epochs, training was automatically terminated. The best network parameters and associated performance metrics were recorded for later evaluation.

In order to improve convergence robustness, a plateau-detection strategy was implemented. In case the validation loss stagnated, for waitCount>20 and $L_{val}>0.35$, all learning rate multipliers were reduced by 50%, i.e., $\eta \leftarrow 1/2\ \eta$, allowing for finer weight updates and to escape possible local minima Prechelt [33]. When the validation loss had reached the “good” region, $L_{val} \leq 0.30L$ and remained stable for a predefined window, the hybrid multi-objective loss function was replaced by an uncertainty-weighted formulation. In this regime, each task-specific loss $L_i$ was automatically scaled by its predictive uncertainty $\sigma_i$:

$$L_{\text{total}} = \sum_i \frac{1}{2\sigma_i^2} L_i + \log \sigma_i \quad (29)$$

As presented in the work of Kendall and Gal [34] This adaptive weighting scheme gives higher weights to tasks having lower predictive confidence while regularizing those with higher uncertainty, thus balancing multi-task optimization.

Gradient norms were also periodically checked during training for the stability of the optimization process. The Frobenius norm of each parameter group's gradients was computed as in equation (30) and visualized every 10 epochs for the backbone and both task heads. This ensured that learning dynamics remained inside expected ranges in terms of magnitudes and that no parameter sub-group dominated updates.

$$\|\nabla_\theta L\|_F = \sqrt{\sum_{i,j} \left(\partial L / \partial \theta_{ij}\right)^2} \quad (30)$$

**4. Network performance Evaluation**

**4.1 Quantitative Evaluation Metrics**

The best-performing model corresponding to the lowest validation loss was stored after training for evaluation. The entire dataset was assessed using several quantitative performance metrics such as the mean absolute percentage error (MAPE) and the correlation coefficient of determination ($R^2$) defined respectively as Chicco et al. [35]:

$$\text{MAPE} = \frac{100}{N} \sum_i \left| \frac{y_i - \hat{y}_i}{y_i} \right| \quad (31)$$

$$R^2 = 1 - \frac{\sum_i (y_i - \hat{y}_i)^2}{\sum_i (y_i - \bar{y})^2} \quad (32)$$

Also the Pearson correlation (Benesty et al. [32]) was computed between predictions and ground-truth targets to evaluate linear correspondence:

$$r = \frac{\sum_i (y_i - \bar{y})(\hat{y}_i - \bar{\hat{y}})}{\sqrt{\sum_i (y_i - \bar{y})^2 \sum_i (\hat{y}_i - \bar{\hat{y}})^2}} = \frac{\text{cov}(y, \hat{y})}{\sigma_y \sigma_{\hat{y}}} \quad (33)$$

where $N$= number of samples, $\bar{y}$= mean of y, $\sigma_y, \sigma_{\hat{y}}$= standard deviations.

These combined metrics provide a comprehensive evaluation of predictive accuracy, calibration, and correlation for both the density and auxiliary outputs. Scatter plots visualizing prediction quality are also plotted for evaluation.

Table 2 represents the metrics performance in terms of $R^2$, MAPE, and Pearson correlation, for the predicted versus true values of: density in $g.cm^{-3}$, specific compression modulus ($MPa/g.cm^{-3}$), specific yield stress ($MPa/g.cm^{-3}$) and specific compression strength ($MPa/g.cm^{-3}$).

**Table 2: Performance metric of the density and mechanical properties predictions averaged for all samples**

| Property | $R^2$ | AVERAGE MAPE (%) | Correlation (r) | Interpretation |
|---|---|---|---|---|
| **Density** | 0.8505 | 5.565 | 0.922 | Excellent predictive agreement with minimal error, indicating accurate modeling of density variations. |
| **Specific Compression Modulus** | 0.897 | 8.738 | 0.947 | Very good prediction performance: slightly higher deviation attributed to broader data spread in stiffness measurements |
| **Specific Yield Stress** | 0.888 | 9.127 | 0.943 | Strong linear correlation and reliable predictive behavior, though minor dispersion reflects inherent experimental variability. |
| **Specific Compression Strength** | 0.914 | 6.163 | 0.956 | Highest predictive fidelity among all properties; excellent consistency and robust regression performance. |

Figure 5 represents the predicted vs. true values scatter plots where it can be seen how the predicted align perfectly with the true values showcasing the robustness of the proposed artificially intelligent method to predict properties.

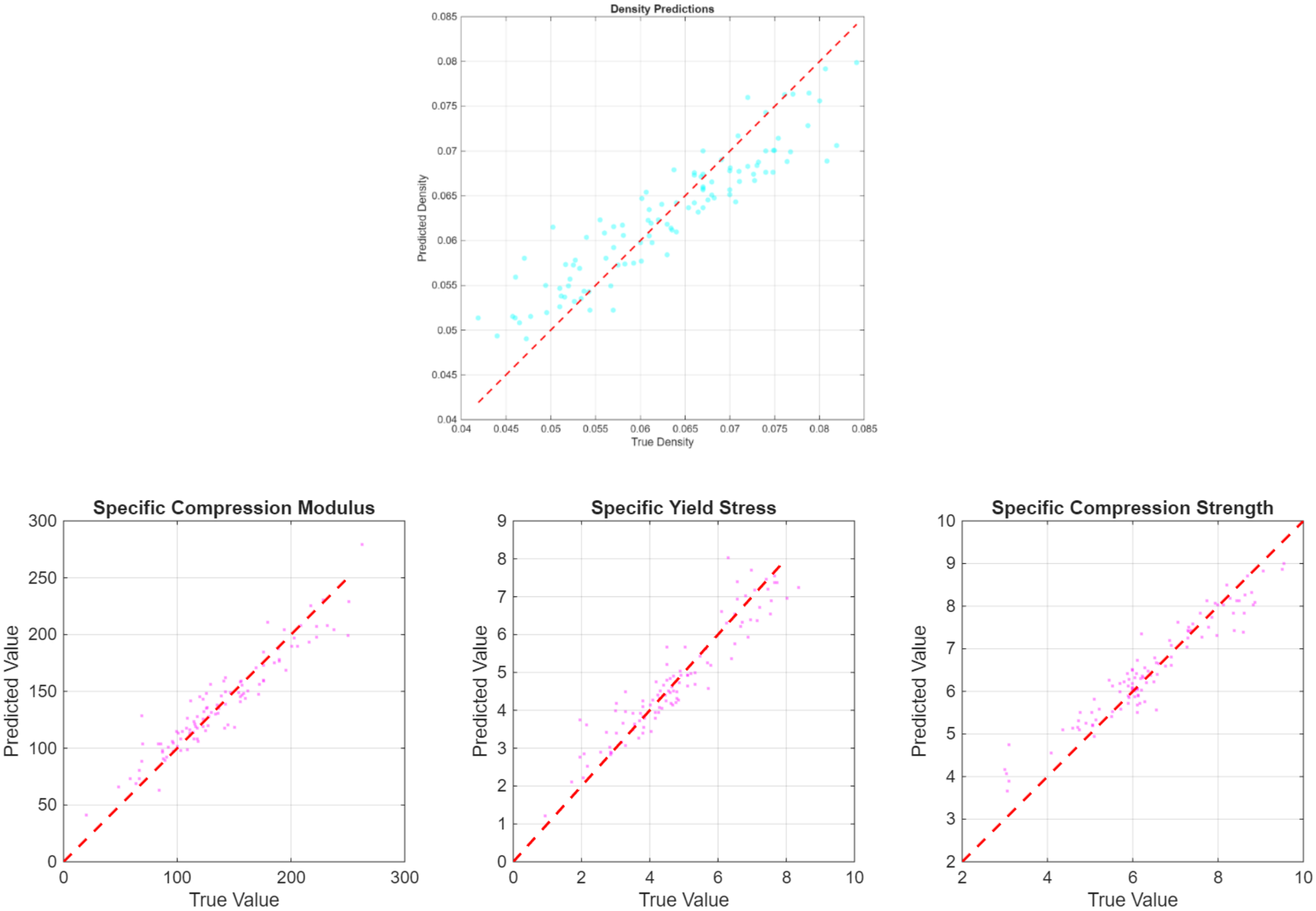

**Figure 5: Predicted versus True values of : (top) the density in $g.cm^{-3}$, (bottom) Specific compression modulus ($MPa/g.cm^3$), Specific yield stress ($MPa/g.cm^3$) and Specific compression strength ($MPa/g.cm^{-3}$)**

According to Table 2 and Figure 5, the predictive performance of the proposed convolutional network exhibits a strong agreement between the predicted and experimental mechanical properties of the lignin-modified polyurethane foams. The model achieved coefficients of determination ($R^2$ = 0.850–0.91) across all four targets with correlation coefficients ($r > 0.92$) confirming excellent linear consistency between predicted and true values.

The MAPE for all the predicted properties did not exceed 9.12 %, thus confirming strong generalization capability and minimal bias of the model. Among the target outputs, the compression strength showed the highest predictive reliability ($R^2$ = 0.914; MAPE = 6.16 %; r = 0.956), closely followed by the compression modulus ($R^2$ = 0.897; MAPE = 8.74 %; r = 0.947). Yield stress and density predictions also showed strong agreement with the experimental data: $R^2$ = 0.888 and 0.851, MAPE = 9.13 % and 5.57 %, and r = 0.943 and 0.922, respectively, indicating that the network can grasp complex structure-property relationships across heterogeneous foam morphologies.

Slight changes in the stiffness-related properties result mainly from some local variations in pore geometry as well as cell-wall anisotropy, which are not completely resolved in 2D microscopy inputs. The overall performance of this method, however, indicates that the developed CNN is a physically consistent, data-efficient surrogate model that can predict the important mechanical responses from the microstructure with very little overfitting and strongly representative cross-sample robustness.

### 4.2 Model Interpretability and Robustness Analysis Using Grad-CAM

#### 4.2.1 Visualization of Pore Morphology and Feature Attribution using Grad-CAM

Gradient-weighted Class Activation Mapping (Grad-CAM) proposed in Selvaraju et al., [36] was utilized to improve interpretability, and it was employed to visualize the regions within each SEM image that most strongly influenced the network's predictions.

Several recent studies Shin et al., [16]; Zhao et al., [37],[38] have utilized this approach for identifying critical microstructural features. For that reason, the dual-head CNN developed in this study trained to predict both density and mechanical properties was further analyzed using Grad-CAM applied independently to each output head.

For each prediction, Grad-CAM generated heat maps that highlight the most influential microstructural regions in warmer colors to provide insight into which morphological features the network uses to infer material behavior. Figure 6 represents Grad-CAM heat maps of Control, LHO, KL, LHO-MD, LHO-O, DCA-D, and DCA PU foams that highlight regions which drive the CNNs predictions: highly activated areas appear consistently along cell walls, struts, and junctions, whereas pore voids always remain inactive, showing that the network focuses on the load-bearing skeleton of the porous structure rather than on superficial textures.

The density head presents localizations of activation around thicker struts and hence reduced pore volume, suggesting that the model interprets density variations in terms of cell-wall continuity and local concentration of mass.

By contrast, the mechanical-property heads, Ec, σy, and σc, exhibit broader activation fields spanning interconnected junction networks and aligned pore chains features that are physically responsible for the stress transfer and resistance to deformation.

Across all outputs, the network effectively internalized key microstructural descriptors such as pore connectivity, anisotropy, and cell-aspect ratio, known to govern both stiffness and strength in open-cell foams. Analysis of the convolutional hierarchy reveals that shallow layers capture fine-scale pore boundaries and surface roughness, whereas deep layers abstract these into global measures of cell-size distribution and orientation. The correspondence of the red activation zones with dense strut regions serves as confirmation that physically meaningful geometry-property relations have been learned by the CNN, rather than spurious image correlations.

In conclusion, the Grad-CAM analysis serves to further corroborate the dual-head CNN as a physically interpretable and reliable model in predicting mechanical performance from porous microstructures, thus enhancing its suitability for the data-driven study of structure-property relationships within lignin-based polyurethane foams.

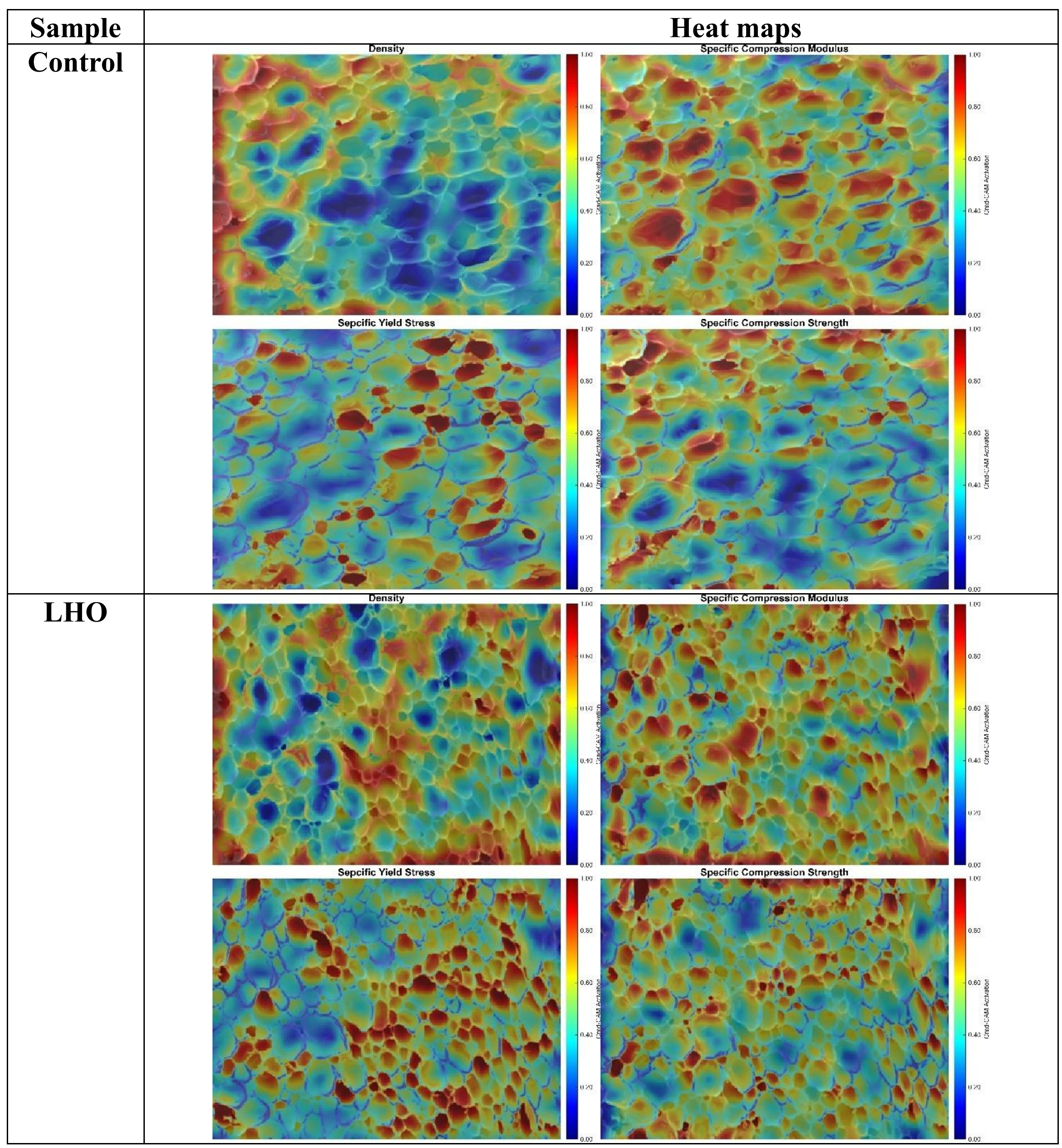

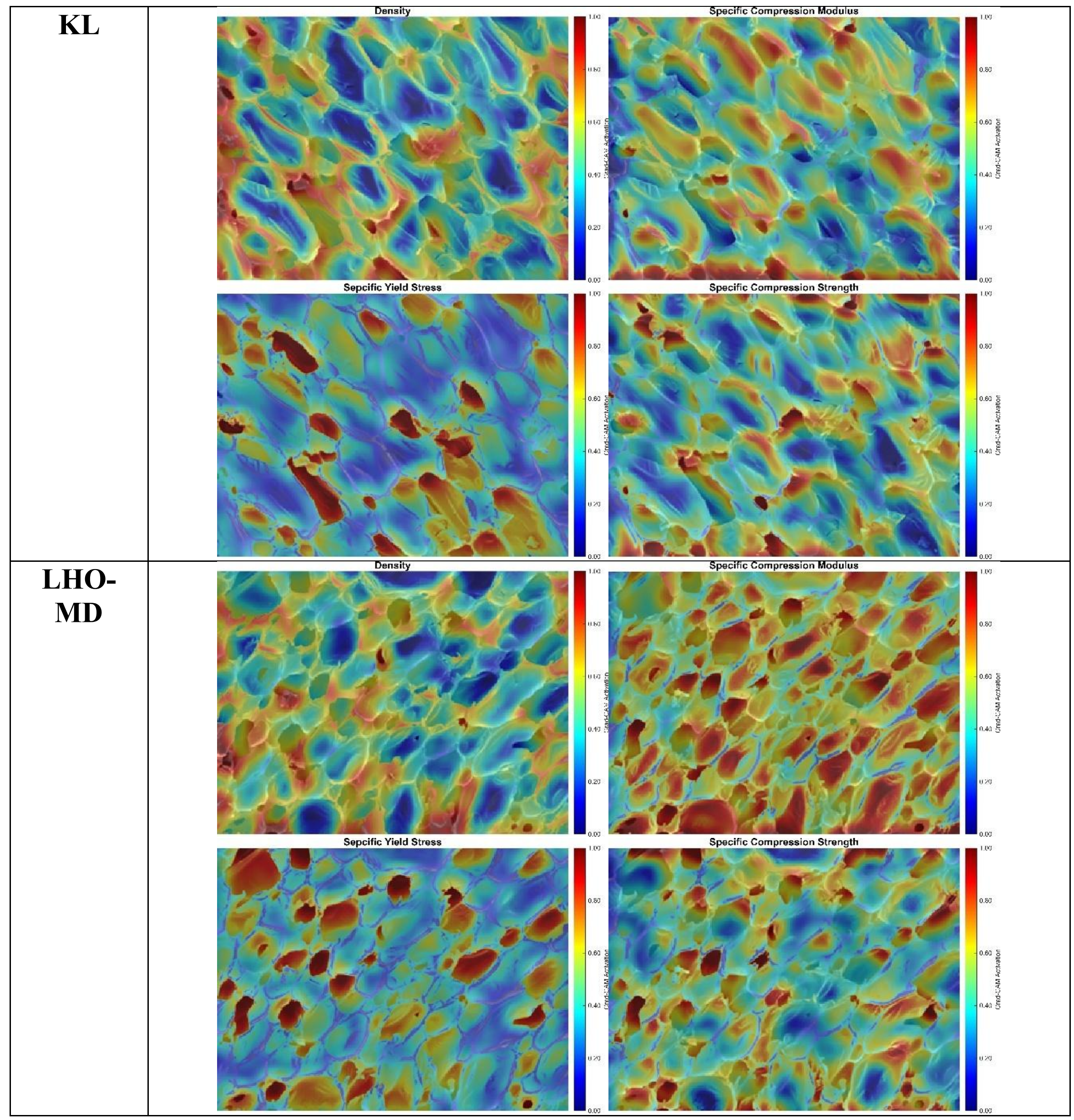
KL
LHO-MD
Density
Specific Compression Modulus
Sepcific Yield Stress
Specific Compression Strength

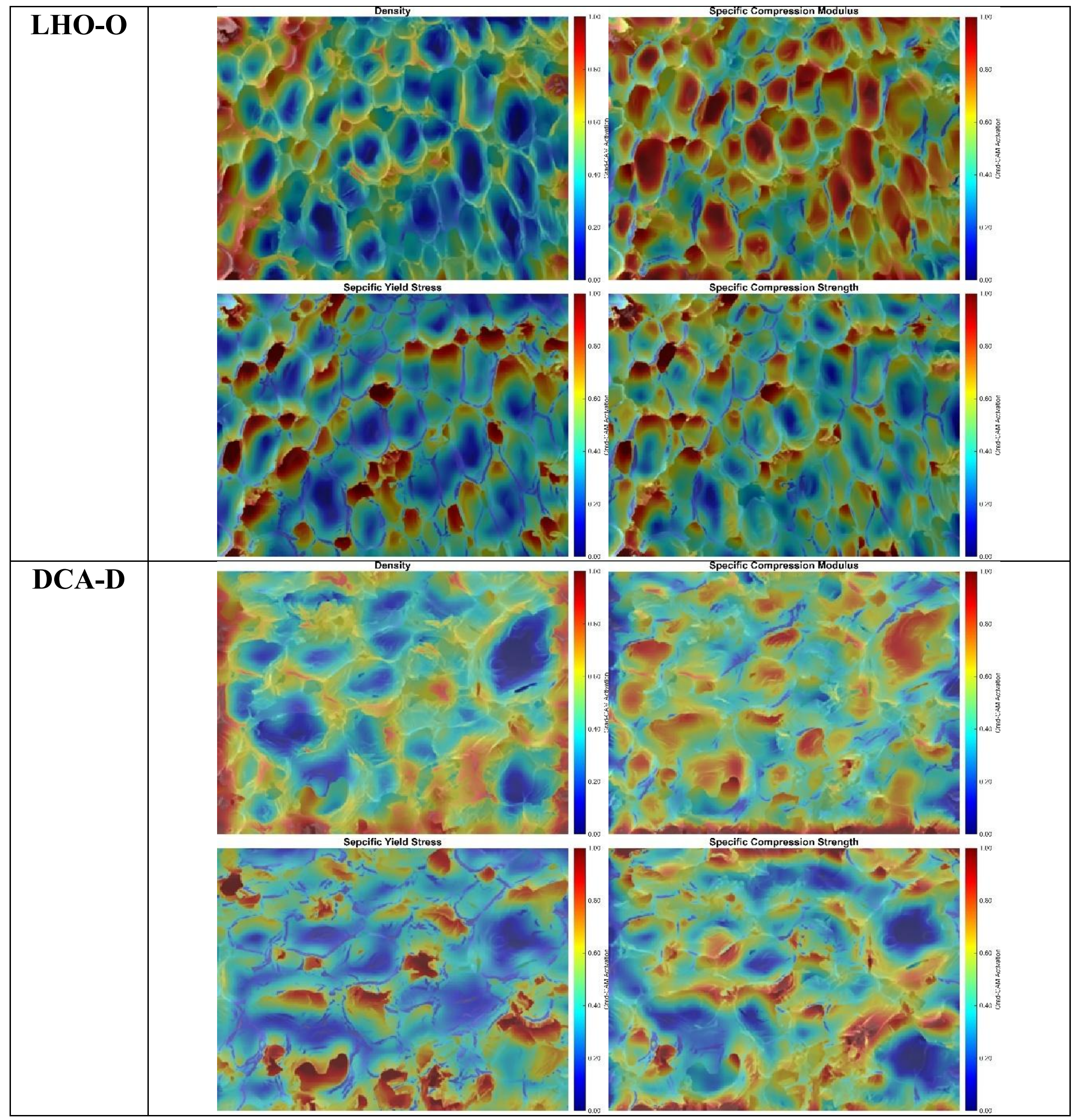
LHO-O
Density
Specific Compression Modulus
Sepcific Yield Stress
Specific Compression Strength
DCA-D
Density
Specific Compression Modulus
Sepcific Yield Stress
Specific Compression Strength

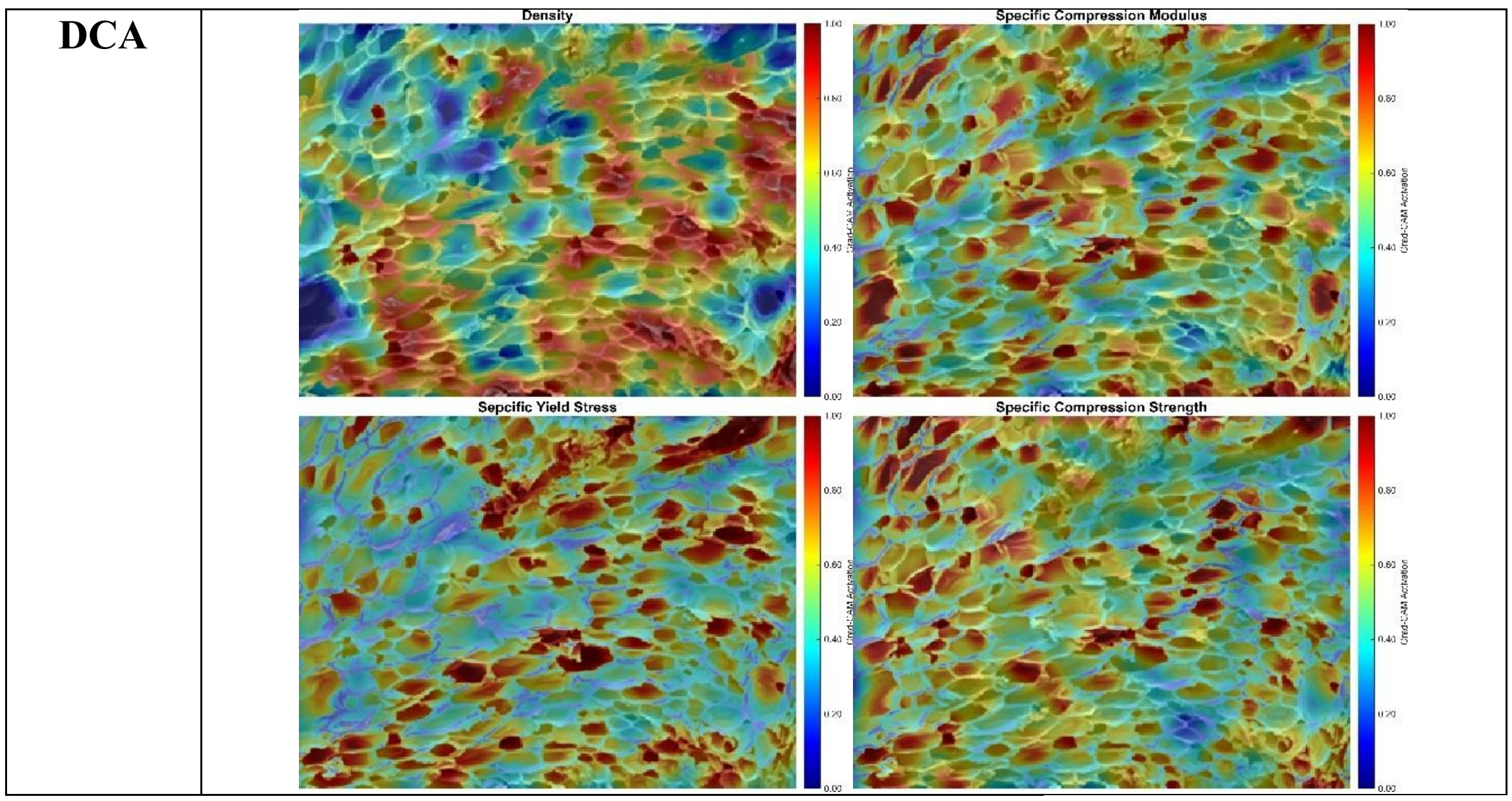

**Figure 6: Heat maps for the selected images in figure 1 (PU foams prepared using the different lignin-based polyols at a loading of 25 wt % ) for all the properties ( density ($g\cdot cm^{-3}$), specific compression modulus (MPa/( $g\cdot cm^{-3}$)), specific yield strength (MPa/( $g\cdot cm^{-3}$)) and specific compression strength (MPa/( $g\cdot cm^{-3}$))) calculated**

### 4.2.2 Interpretation of Grad-CAM Visualizations and associated quantitative results

Despite that Grad-CAM is not a quantitative performance metric, it helps as a qualitative interpretation tool that reveals the microstructural regions most helping the model's predictions. To complement this, quantitative comparisons between true predicted values were reported.

A comparison between experimental and CNN-predicted density, specific compression modulus, specific yield strength, and specific compression strength for the same foam formulations at 25 wt.% lignin associated with the samples shown in figure 1 and 6 is presented in table 3. Small relative errors, typically 1-5%, max below 9%, and correct stiffness and strength ranking across materials preserved provide quantitative evidence that the features highlighted in the Grad-CAM heat maps translate into accurate, physically consistent predictions for porous media. Quantitative results in Table 3 confirm the high predictive fidelity of the model.

Figure 6 representing Grad-CAM visualizations combined with the quantitative results of Table 3 demonstrate that the trained dual-head CNN learned to associate physically meaningful microstructural regions with the corresponding macroscopic properties of the lignin-based PU foams. For all the output heads, density, specific compression modulus, specific yield stress, and specific compression strength, the model shows strong activation intensities along the cell walls, strut junctions, and pore edges, while the inner voids remain largely inactive.

This spatial activation pattern indicates that the CNN identified the load-bearing skeleton of the foam as the principal contributor to its mechanical response, as would be expected from established structure-property relationships for open-cell polymers. Values for the specific compression modulus, the yield strength, the compression strength, and the density predicted for the Control sample are very close to the experimental ones (117.52 → 117.29 MPa/( $g \cdot cm^{-3}$), 4.18 → 4.25 MPa/($g \cdot cm^{-3}$), 5.99 → 6.07 MPa/($g \cdot cm^{-3}$), and 0.067 → 0.066 $g \cdot cm^{-3}$, respectively), with errors below 2%. Similarly, in the DCA formulation, the overall deviation is lowest (< 2 %), while density and modulus trends have been reproduced for highly isotropic microstructures, e.g., 155.62 → 158.67 MPa/($g \cdot cm^{-3}$) and 4.60 → 5.00 MPa/($g \cdot cm^{-3}$), for modulus and strength, respectively. This consistency indicates network robustness regarding the capture of subtle structure–property correlations even under limited morphological variability. Slightly higher deviations were recorded for LHO-MD and LHO-O foams, where the density and stiffness errors attained 3–8%.

These differences likely arise from greater microstructural heterogeneity, including elongated pores and anisotropic cell orientations, which complicate feature abstraction at the convolutional level. Nonetheless, even for such complicated morphologies, the CNN continued to make precise strength predictions and correctly rank the material by stiffness. It had, therefore, successfully generalized beyond the training distribution.

**Table 3: Predicted vs. Actual properties of density ($g \cdot cm^{-3}$), specific compression modulus (MPa/($g \cdot cm^{-3}$)), specific yield strength (MPa/($g \cdot cm^{-3}$)) and specific compression strength (MPa/( $g \cdot cm^{-3}$)) of the PU foams prepared using the different lignin-based polyols at a loading of 25 wt.% of the polyol component**

| Material | Property | Actual | Predicted | Absolute Error | % Error (%) |
|---|---|---|---|---|---|
| **Control (CF)** | Density ($g \cdot cm^{-3}$) | 0.067 | 0.0660 | 0.0010 | **1.49** |
| | Specific Compression Modulus (MPa/($g \cdot cm^3$)) | 117.515 | 117.291 | 0.224 | **0.19** |
| | Specific Yield Strength (MPa/($g \cdot cm^3$)) | 4.182 | 4.255 | 0.073 | **1.73** |
| | Specific Compression Strength (MPa/($g \cdot cm^3$)) | 5.995 | 6.074 | 0.079 | **1.32** |
| **LHO** | Density ($g \cdot cm^{-3}$) | 0.063 | 0.0584 | 0.0046 | **7.30** |
| | Specific Compression Modulus (MPa/($g \cdot cm^3$)) | 115.629 | 120.953 | 5.324 | **4.60** |
| | Specific Yield Strength (MPa/($g \cdot cm^3$)) | 4.224 | 4.395 | 0.171 | **4.05** |
| | Specific Compression Strength (MPa/($g \cdot cm^3$)) | 5.900 | 5.892 | 0.008 | **0.14** |
| **KL** | Density ($g \cdot cm^{-3}$) | 0.04945 | 0.04701 | 0.00245 | **4.95** |
| | Specific Compression Modulus (MPa/($g \cdot cm^3$)) | 138.611 | 145.905 | 7.294 | **5.26** |
| | Specific Yield Strength (MPa/($g \cdot cm^3$)) | 4.308 | 4.500 | 0.192 | **4.45** |
| | Specific Compression Strength (MPa/($g \cdot cm^3$)) | 6.411 | 6.380 | 0.031 | **0.48** |
| **LHO-MD** | Density ($g \cdot cm^{-3}$) | 0.06340 | 0.06137 | 0.00203 | **3.21** |
| | Specific Compression Modulus (MPa/($g \cdot cm^3$)) | 202.499 | 197.239 | 5.260 | **2.60** |
| | Specific Yield Strength (MPa/($g \cdot cm^3$)) | 7.525 | 7.542 | 0.017 | **0.23** |

| | Specific Compression Strength (MPa/(g·$cm^3$)) | 7.317 | 7.496 | 0.179 | **2.44** |
|---|---|---|---|---|---|
| **LHO-O** | Density (g·$cm^{-3}$) | 0.0606 | 0.0654 | 0.0048 | **7.88** |
| | Specific Compression Modulus (MPa/(g·$cm^3$)) | 176.022 | 172.851 | 3.171 | **1.80** |
| | Specific Yield Strength (MPa/(g·$cm^3$)) | 4.674 | 4.613 | 0.061 | **1.31** |
| | Specific Compression Strength (MPa/(g·$cm^3$)) | 7.731 | 8.136 | 0.405 | **5.23** |
| **DCA-D** | Density (g·$cm^{-3}$) | 0.06646 | 0.06317 | 0.00329 | **4.96** |
| | Specific Compression Modulus (MPa/(g·$cm^3$)) | 87.979 | 90.858 | 2.879 | **3.27** |
| | Specific Yield Strength (MPa/(g·$cm^3$)) | 2.885 | 2.870 | 0.015 | **0.54** |
| | Specific Compression Strength (MPa/(g·$cm^3$)) | 3.033 | 3.073 | 0.040 | **1.32** |
| **DCA** | Density (g·$cm^{-3}$) | 0.04576 | 0.04515 | 0.00061 | **1.33** |
| | Specific Compression Modulus (MPa/(g·$cm^3$)) | 155.621 | 158.667 | 3.046 | **1.96** |
| | Specific Yield Strength (MPa/(g·$cm^3$)) | 6.115 | 6.602 | 0.487 | **7.96** |
| | Specific Compression Strength (MPa/(g·$cm^3$)) | 4.595 | 5.003 | 0.408 | **8.87** |

Looking at the learned feature activations provided by Grad-CAM provided, it gave the model an interpretable bridge between statistical learning and physical mechanisms. Coupling Grad-CAM results in figure 6 with their corresponding quantitative calculation in table 3 proves that the dual-head CNN does not rely on superficial image statistics but rather captures true morphological determinants of mechanical behavior, This demonstrates the proposed architecture as an Artificially intelligent framework for microstructure-driven materials deigns.
To sum up, the model achieved consistent, accurate, physics-aligned predictions across diver lignin foams, with most errors confined between 1-5%. This demonstrates the robustness of the proposed network, however modest deviations highlight the opportunities to improve the model through microstructural-specific data augmentation or hybrid physics informed layers as future enhancements.

## Conclusion

A novel dual-head CNN was developed and validated in this study to quantitatively predict properties of lignin-modified PU rigid foams from SEM images. A shared convolutional layer with three convolution pooling stages constructed the backbone of the network, and a fully connected shared layer that branched into two specialized heads for density and the mechanical property (specific compressive modulus, yield stress and compressive strength) predictions. The model demonstrated excellent performance for all outputs, with $R^2$ ranging from 0.85 to 0.91, Pearson r above 0.92 and MAPE lower than 9.12%, proving its robustness, low bias and strong generalization capability over chemically and morphologically diverse foams. This proposed network was further demonstrated that it learned physically meaningful microstructural mechanical correlation using Grad-CAM visualizations where high activation intensities were found to be consistently localized along load-bearing struts, cell-wall junctions and pore edges, while pore voids remained largely inactive. Numerical agreement between predicted versus measured properties agrees with the physical consistency of the learned representations. This work demonstrated how image-based neural networks can constitute a reliable, non-destructive surrogate for mechanical testing while retaining physical interpretability which helps provide a transparent and scalable framework that bridges morphological imaging with mechanical prediction, thus providing a data-driven route toward the design and optimization of sustainable bio-based PU foams.

## ACKNOWLEDGMENTS

We would like to thank Dr. Lloyd Donaldson for his assistance with SEM measurements. We would also like to acknowledge Maxime Barbier, Regis Risani and Ross Anderson for their support with compression testing. We would like to thank Dr. Kirk Torr for his unwavering help and support that made this work possible.

## STATEMENTS AND DECLARATIONS

The authors have no conflicts of interest to declare.

## DATA AVAILABILITY STATEMENT

The datasets generated and/or analyzed during the current study are not publicly available due to confidentiality and intellectual property restrictions, but are available from the corresponding author on reasonable request.

## AUTHORS CONTRIBUTIONS

All authors contributed to the work in the manuscript.